\documentclass[english,3p]{elsarticle}
\usepackage{lmodern}
\usepackage{lmodern}
\usepackage[T1]{fontenc}
\usepackage[latin9]{inputenc}
\usepackage{xcolor}
\usepackage{pdfcolmk}
\definecolor{document_fontcolor}{rgb}{0, 0, 0}
\color{document_fontcolor}
\usepackage{babel}
\usepackage{amsmath}
\usepackage{amssymb}
\usepackage{graphicx}
\PassOptionsToPackage{normalem}{ulem}
\usepackage{ulem}
\usepackage[unicode=true,
 bookmarks=true,bookmarksnumbered=false,bookmarksopen=false,
 breaklinks=false,pdfborder={0 0 1},backref=false,colorlinks=true]
 {hyperref}
\hypersetup{
 pdfborderstyle=}

\makeatletter

\providecommand{\tabularnewline}{\\}
\providecolor{lyxadded}{rgb}{1,0,0}
\providecolor{lyxdeleted}{rgb}{0.33,0.67,1}
\DeclareRobustCommand{\lyxadded}[3]{{\texorpdfstring{\color{lyxadded}{}}{}#3}}
\DeclareRobustCommand{\lyxdeleted}[3]{{\texorpdfstring{\color{lyxdeleted}\lyxsout{#3}}{}}}
\DeclareRobustCommand{\lyxsout}[1]{\ifx\\#1\else\sout{#1}\fi}
\usepackage{tikz}
\usetikzlibrary{calc}
\newcommand{\lyxmathsout}[1]{%
  \tikz[baseline=(math.base)]{
    \node[inner sep=0pt,outer sep=0pt](math){#1};
    \draw($(math.south west)+(2em,.5em)$)--($(math.north east)-(2em,.5em)$);
  }
}

\usepackage{babel}
\usepackage{array}
\usepackage{bm}
\usepackage{multirow}
\usepackage{braket}
\usepackage{enumitem}
\usepackage{float}
\usepackage{extarrows}
\usepackage[numbers]{natbib}
\biboptions{sort&compress}

\providecommand{\tabularnewline}{\\}

\providecommand{\doi}[1]{%
	\begingroup
	\let\bibinfo\@secondoftwo
	\urlstyle{rm}%
	\href{http://dx.doi.org/#1}{%
		\discretionary{}{}{}%
		\nolinkurl{#1}%
	}%
	\endgroup
}

\usepackage{xcolor}
\usepackage{textcomp}
\usepackage{hyphenat}
\usepackage{makecell}

\usepackage{listings}
\usepackage{xcolor}
\usepackage{mathematica}
\definecolor{darkred}{rgb}{0.5 0 0}
\definecolor{darkgreen}{rgb}{0.5 .5 0}
\definecolor{darkblue}{rgb}{0 0 .5}

\usepackage{listings}

\makeatother

\usepackage{listings}

\begin{document}

\title{MagneticTB: A package for tight-binding model of magnetic and non-magnetic materials}


\author[buct]{Zeying Zhang}

\author[bit1,bit2]{Zhi-Ming Yu}

\author[bit1,bit2]{Gui-Bin Liu\corref{cor}}

\cortext[cor]{Corresponding author}

\ead{gbliu@bit.edu.cn}

\author[bit1,bit2]{Yugui Yao\corref{cor}}


\ead{ygyao@bit.edu.cn}


\address[buct]{College of Mathematics and Physics, Beijing University of Chemical
Technology, Beijing 100029, China}


\address[bit1]{Centre for Quantum Physics, Key Laboratory of Advanced Optoelectronic Quantum Architecture and Measurement (MOE), School of Physics, Beijing Institute of Technology, Beijing, 100081, China}

\address[bit2]{Beijing Key Lab of Nanophotonics \& Ultrafine Optoelectronic Systems, School of Physics, Beijing Institute of Technology, Beijing, 100081, China}

\date{\today}
\begin{abstract}
We present a Mathematica program package MagneticTB, which can generate
the tight-binding model for arbitrary magnetic space group. The only
input parameters in MagneticTB are the (magnetic) space group number
and the orbital information in each Wyckoff positions. Some useful
functions including getting the matrix expression for symmetry operators,
manipulating the energy band structure by parameters and interfacing
with other software are also developed. 
MagneticTB can help to investigate the physical properties in both magnetic and non-magnetic
system, especially for topological properties.
\begin{keyword}
Tight-binding method, Representation theory, Magnetic space group,
Mathematica 
\end{keyword}

\textbf{Program summary}

Program title: MagneticTB


Licensing provisions: GNU General Public Licence 3.0

Programming language: Mathematica

External routines/libraries used: ISOTROPY (iso.byu.edu)

Developer's repository link: https://github.com/zhangzeyingvv/MagneticTB

Nature of problem: Construct the symmetry adopted tight-binding model
for the system with arbitrary magnetic space group.
\end{abstract}
\maketitle


\section{Introduction}

Tight-binding method is a powerful tool to investigate the novel properties
in condensed mater physics
\citep{wallace_band_1947,lowdin_nonorthogonality_1950,slater_simplified_1954,goringe_tight-binding_1997,wieder_wallpaper_2018}.
Compared with first-principles method, tight-binding method can greatly
simplify calculations. Moreover, after considering (magnetic) space
group symmetry, the tight-binding model can give more reliable results.
For example, in topological materials, symmetry plays
an important role to protect the topological properties, such as $Z_{2}$
topological insulator protected by time reversal symmetry \citep{kane_z2_2005},
topological crystalline insulators and topological nodal semimetals
protected by space group symmetries \citep{fu_topological_2011,burkov_topological_2011}
and 
magnetic topological crystalline insulator protected by magnetic
space group symmetries, i.e. the combination of space group operations
and time reversal \citep{fang_topological_2013,liu_antiferromagnetic_2013}.

At present, a lot of researchers using symmetry adopted tight-binding
model to investigate physical properties of electronic system \citep{egorov_consistent_1968,kuznetsov_symmetry_1978,ku_insulating_2002,liu_three-band_2013,wieder_spin-orbit_2016,wang_hourglass_2016,song__2017,zhang_magnetization-direction_2018,gresch_automated_2018,koshino_maximally_2018,yu_circumventing_2019}.
However, most of the software packages are mainly focused on the tight-binding
model for non-magnetic materials \citep{di_martino_high_1999,groth_kwant_2014,supka_aflow_2017,nakhaee_tight-binding_2020,klymenko_nanonet_2021}
and only a few of them can be used to construct the symmetry adopted
tight-binding model automatically. For the first-principles level,
Wannier90 can generate the Wannier-tight-binding model by interfacing
with first-principles software, but the symmetry adopted Wannier function
can not be applied to other first-principles software (such as VASP,
ABINIT) except Quantum-Espresso \citep{mostofi_wannier90:_2008,giannozzi_quantum_2009,kresse_efficient_1996,gonze_abinit_2009}.
FPLO can generate symmetry adopted tight-binding model with proper
parameters for given structures \citep{koepernik_full-potential_1999}.
Meanwhile, both Wannier90 and FPLO do not support magnetic symmetry.
For the model level, GTPack, Qsymm and MathemaTB can generate the
tight-binding model with space group symmetry but does
not support magnetic symmetry directly \citep{hergert_group_2018,varjas_qsymm_2018,jacobse_mathematb_2019}.
WannierTools can do the symmetrization of non-magnetic tight-binding
model but cannot generate the tight-binding model by itself \citep{wu_wanniertools:_2018}.

Therefore, it is necessary to develop a package which can construct
the tight-binding model with magnetic space group symmetry automatically.
Here we introduce a software package: MagneticTB, a tool for generating
the tight-binding model for the system with arbitrary magnetic space
group. The required input information of this package is only the
magnetic space group number and the orbital information in each Wyckoff
positions. It can help to investigate the physical properties of the
given symmetry. We also present some useful functions including get
the matrix expression for symmetry operators, manipulate the band
structure by parameters and interface with other software.

This paper is organized as follows. In Sec.~\ref{sec:Methods}, we
give an introduction of symmetry adopted tight-binding methods. In
Sec.~\ref{sec:Capabilities}, The usage of MagneticTB are given,
including how to install and run MagneticTB. In Sec.~\ref{sec:example},
we give three concrete examples, such examples show the specific capabilities
of the MagneticTB. Finally, conclusions are given. 

\section{Symmetry adopted tight-binding method}

\label{sec:Methods}

In periodic system, the bases
of tight-binding model can be written as Bloch sum \citep{egorov_consistent_1968}
\begin{equation}
\psi_{lm\boldsymbol{k}}^{n}(\boldsymbol{r})=\frac{1}{\sqrt{N}}\sum_{\boldsymbol{R}_{j}}e^{i\boldsymbol{k}\cdot(\boldsymbol{R}_{j}+\boldsymbol{d}_{l}^{n})}\varphi_{m}^{n}(\boldsymbol{r}
-\boldsymbol{R}_{j}-\boldsymbol{d}_{l}^{n})\label{eq:blcochsum}
\end{equation}
where $N$ is the number of unit cells in the crystal, $\boldsymbol{R}_{j}$
is the translation vector of the Bravais lattice, $\boldsymbol{d}_{l}^{n}$
is the position of $n$-th Wyckoff position's $l$-th atom in the
unit cell (for each $Q$  and $\boldsymbol{d}_{l}^{n} $ there
exists one and only one pair of $\boldsymbol{d}_{l'}^{n}$ and $R'$ which satisfies 
$Q\boldsymbol{d}_{l}^{n}=\boldsymbol{d}_{l'}^{n}+ R'$ where  $Q$ is arbitrary group element in the (magnetic)
space group and $R'$ is a lattice vector),
$\varphi_{m}^{n}(\boldsymbol{r})$
is the $m$-th atomic orbital
basis for
position $\boldsymbol{d}_{l}^{n}$ but
located at coordinate origin. Then Eq.~(\ref{eq:blcochsum}) satisfies
the Bloch theorem $\psi_{lm\boldsymbol{k}}^{n}(\boldsymbol{r}+\boldsymbol{R}_{j})=e^{i\boldsymbol{k}\cdot\boldsymbol{R}_{j}}\psi_{lm\boldsymbol{k}}^{n}(\boldsymbol{r})$.
The tight-binding Hamiltonian can be written as: 
\begin{equation}
\begin{split}H_{lml'm'}^{nn'}(\boldsymbol{k})= & \sum_{\boldsymbol{R}_{j}}e^{i\boldsymbol{k}\cdot(\boldsymbol{R}_{j}+\boldsymbol{d}_{l'}^{n'}-\boldsymbol{d}_{l}^{n})}E_{mm'}(\boldsymbol{d}_{l}^{n},\boldsymbol{R}_{j}+\boldsymbol{d}_{l'}^{n'})\\
E_{mm'}(\boldsymbol{d}_{j}^{n},\boldsymbol{R}_{j}+\boldsymbol{d}_{l'}^{n'})= & \braket{\varphi_{m}^{n}(\boldsymbol{r}-\boldsymbol{d}_{l}^{n})|\hat{H}|\varphi_{m'}^{n'}(\boldsymbol{r}-\boldsymbol{d}_{l'}^{n'}-\boldsymbol{R}_{j})}
\end{split}
\label{eq:covI}
\end{equation}
for simplicity, we rewrite the atomic orbitals
in vector form: $\varPhi^{n}(\boldsymbol{r}-\boldsymbol{R}_{j}-\boldsymbol{d}_{l}^{n})=\{\varphi_{m}^{n}(\boldsymbol{r}-\boldsymbol{R}_{j}-\boldsymbol{d}_{l}^{n})\},(m=1,...,M_{n})$.
Then $E_{mm'}(\boldsymbol{d}_{j},\boldsymbol{R}_{j}+\boldsymbol{d}_{l'})$
$(m=1,...,M_{n};\,m'=1,\dots,M_{n'})$ form an
$M_{n}\times M_{n'}$ matrix: 
\begin{equation}
E(\boldsymbol{d}_{l}^{n},\boldsymbol{R}_{j}+\boldsymbol{d}_{l'}^{n'})=\braket{\varPhi^{n}(\boldsymbol{r}-\boldsymbol{d}_{l}^{n})|\hat{H}|\varPhi^{n'}(\boldsymbol{r}-\boldsymbol{R}_{j}-\boldsymbol{d}_{l'}^{n'})}
\end{equation}
\begin{figure}[t]
\includegraphics[width=1\textwidth]{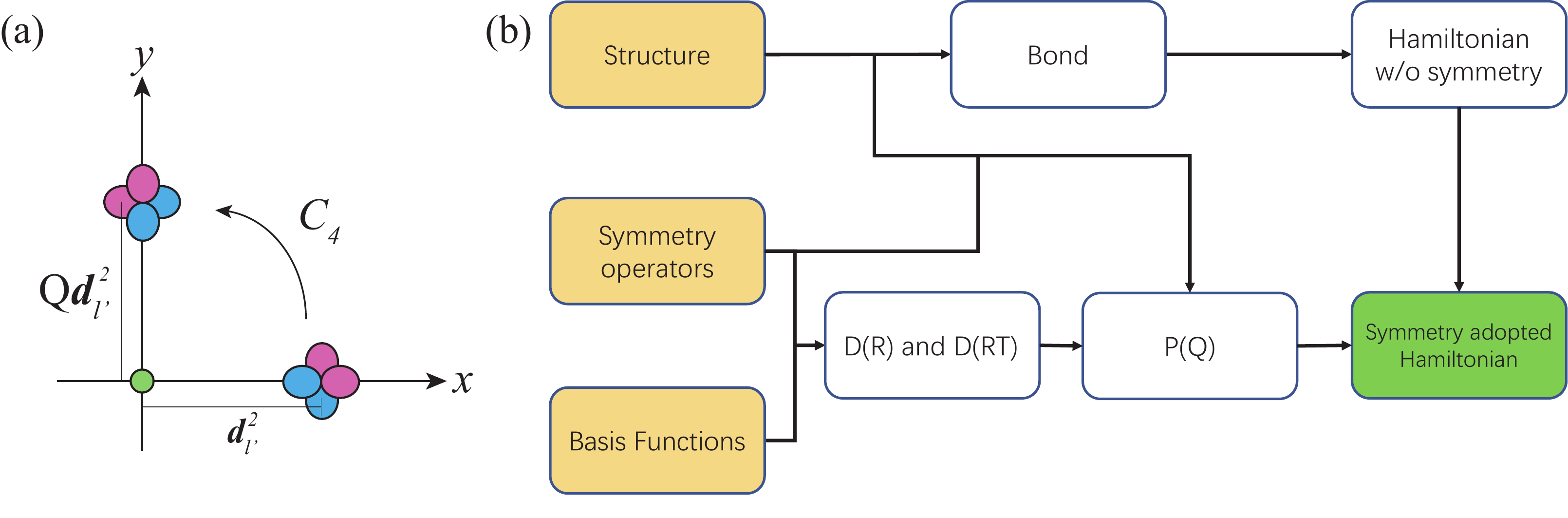} \caption{(a) Sketch of relationship between $E(\boldsymbol{d}_{j}^{n},\boldsymbol{R}_{j}+\boldsymbol{d}_{l'}^{n'})$
and $E(Q\boldsymbol{d}_{j}^{n},Q(\boldsymbol{R}_{j}+\boldsymbol{d}_{l'}^{n'}))$,
in this example we set $Q=C_{4}{\cal T}$, $\varPhi^{1}(\boldsymbol{r})=\{s\}$
locate at $\boldsymbol{d}_{l}^{1}=(0,0)$, $\varPhi^{2}(\boldsymbol{r})=\{p_{x},p_{y}\}$
locate at $\boldsymbol{d}_{l'}^{2}=(\lambda,0)(\lambda\protect\neq0)$.
Then we have $D^{1}(C_{4}{\cal T})=1,D^{2}(C_{4}{\cal T})=-i\sigma_{y}$,
and $E(Q\boldsymbol{d}_{l}^{1},Q\boldsymbol{d}_{l'}^{2})=E^{*}(\boldsymbol{d}_{l}^{1},\boldsymbol{d}_{l'}^{2})\times(-i\sigma_{y})$.
(b) Workflow of MagneticTB.}
\label{fig:fig1} 
\end{figure}

Then
the Hamiltonian can be rewritten as: 
\begin{equation}
H_{ll'}^{nn'}(\boldsymbol{k})=\sum_{\boldsymbol{R}_{j}}e^{i\boldsymbol{k}\cdot(\boldsymbol{R}_{j}+\boldsymbol{d}_{l'}^{n'}-\boldsymbol{d}_{l}^{n})}E(\boldsymbol{d}_{l}^{n},\boldsymbol{R}_{j}+\boldsymbol{d}_{l'}^{n'})
\end{equation}
$E(\boldsymbol{d}_{l}^{n},\boldsymbol{R}_{j}+\boldsymbol{d}_{l'}^{n'})$
is the hopping matrix between $n$-th Wyckoff position's $l$-th atom
to $n'$-th Wyckoff position's $l'$-th atom. When the lattice is
in invariant under some symmetry $E(\boldsymbol{d}_{l}^{n},\boldsymbol{R}_{j}+\boldsymbol{d}_{l'}^{n'})$
may ne not independent
for arbitrary $\boldsymbol{d}_{l}^{n}$ and $\boldsymbol{d}_{l'}^{n'}$.
Fortunately, for symmetry operation
$Q$, group representation theory gives
us explicit expression for the relationship between $E(Q\boldsymbol{d}_{j}^{n},Q(\boldsymbol{R}_{j}+\boldsymbol{d}_{l'}^{n'}))$
and $E(\boldsymbol{d}_{j}^{n},\boldsymbol{R}_{j}+\boldsymbol{d}_{l'}^{n'})$.

For the case that symmetry operation
does not contain the time
reversal ${\cal T}$, i.e. $Q=\{R|\boldsymbol{t}\}$, where $R$
and $\boldsymbol{t}$ are
the rotation and translation
part of $Q$ respectively,
we have 
\begin{equation}
E(Q\boldsymbol{d}_{l}^{n},Q(\boldsymbol{R}_{j}+\boldsymbol{d}_{l'}^{n}))=D^{n}(R)E(\boldsymbol{d}_{l}^{n},\boldsymbol{R}_{l}+\boldsymbol{d}_{l'}^{n'})D^{n'\dagger}(R)\label{eq:hopp}
\end{equation}
For the case that operation $Q$ contains time
reversal symmetry ${\cal T}$,
i.e. $Q=\{R|\boldsymbol{t}\}{\cal T}$,
we have 
\begin{equation}
E(Q\boldsymbol{d}_{l}^{n},Q(\boldsymbol{R}_{j}+\boldsymbol{d}_{l'}^{n}))=D^{n}(R{\cal T})E^{*}(\boldsymbol{d}_{l}^{n},\boldsymbol{R}_{j}+\boldsymbol{d}_{l'}^{n'})D^{n'\dagger}(R{\cal T})\label{eq:hopptri}
\end{equation}
Where $D^{n}(R)(D^{n}(R{\cal T}))$ are the $M_{n}\times M_{n}$ representation
matrices of $R(R{\cal T})$ under
atomic orbital bases
$\varPhi^{n}(\boldsymbol{r})$ (not necessarily irreducible representations),
$E^{*}(\boldsymbol{d}_{l}^{n},\boldsymbol{R}_{j}+\boldsymbol{d}_{l'}^{n'})$
is complex conjugate of $E(\boldsymbol{d}_{l}^{n},\boldsymbol{R}_{j}+\boldsymbol{d}_{l'}^{n'})$
(see Fig.~\ref{fig:fig1} for example). It is clear that for spinless
cases with time reversal symmetry ${\cal T}$, when the basis functions
are real, $D({\cal T})$ is equal to identity matrix, indicating
that $E(\boldsymbol{d}_{l}^{n},\boldsymbol{R}_{j}+\boldsymbol{d}_{l'}^{n'})$
are real matrices.

The next step is to get the analytical expressions of $D^{n}(R)(D^{n}(RA))$.
For a fixed $n$, we don't have to worry about mixing superscripts
of $n$ in $D^{n}(R)$
because transformations can only occur under the same $n$. So we
temporarily use $D(R)$ rather than $D^{n}(R)$ in this step. Consider
the following four cases: 
\begin{enumerate}[label=\roman*.]
\item Spinless system and $Q$ does not contain ${\cal T}$. 
\item Spinless system and $Q$ contain ${\cal T}$. 
\item Spinful system and $Q$ does not contain ${\cal T}$. 
\item Spinful system and $Q$ contain ${\cal T}$. 
\end{enumerate}
In case.~i, $D(R)$ can be obtained simply by solve the linear
equation \citep{cloizeaux_orthogonal_1963} 
\begin{equation}
\hat{R}\varPhi(\boldsymbol{r})=\varPhi(R^{-1}\boldsymbol{r})=\varPhi(\boldsymbol{r})D(R)\label{eq:dr}
\end{equation}
in which $\hat{R}$ is
the function operator for the rotation $R$. 
In case.~ii, we define $\overline{\varPhi}(\boldsymbol{r})=\hat{{\cal T}}\varPhi(\boldsymbol{r})$,
for spinless system ${\cal T}={\cal K}$, hence, $\overline{\varPhi}(\boldsymbol{r})=\varPhi^{*}(\boldsymbol{r})$
and then solve the linear
equation 
\begin{equation}
\hat{R}\hat{{\cal T}}\varPhi(\boldsymbol{r})=\overline{\varPhi}(R^{-1}\boldsymbol{r})=\varPhi(\boldsymbol{r})D(R{\cal T})\label{eq:dra}
\end{equation}
In case.~iii, since the spin matrix is orbital independent we define
the basis function as 
\begin{equation}
\varPhi^{s}(\boldsymbol{r})={\{\varPhi(\boldsymbol{r})\uparrow,\varPhi(\boldsymbol{r})\downarrow\}}
\end{equation}
The
two spinors $\uparrow=(1,0)^{T}$, and $\downarrow=(0,1)^{T}$,and under the
rotation  $\hat{R}$
they are transformed according
to
\begin{equation}
\hat{R}(\uparrow,\downarrow)=(\uparrow,\downarrow)D^{\frac{1}{2}}(R)
\end{equation}
For
proper rotation $R$, $D^{\frac{1}{2}}(R)=\text{exp}(-\frac{1}{2}i\alpha\boldsymbol{n}\cdot\hat{\boldsymbol{\sigma}})$,
where $\alpha$ is the rotation angle of $R$, $\boldsymbol{n}$ is
the unit vector along rotation axis, for improper rotation $S$, $R=IS$,
$I$ is the inversion symmetry, $D^{\frac{1}{2}}(S)=D^{\frac{1}{2}}(R)$
\citep{landau_chapter_1977,liu_spacegroupirep_2020}. Then $D(R)$
can be obtained by solving the linear
equation 
\begin{equation}
\hat{R}\varPhi^{s}(\boldsymbol{r})=\varPhi^{s}(\boldsymbol{r})D(R)
\end{equation}
Case.~(iv) is similar to case.~(ii), the only difference is replace
the time reversal operator ${\cal T}={\cal K}$ by ${\cal T}=i\hat{\sigma}_{y}{\cal K}$
and consider the spin rotation matrices. The above four cases cover
all the possibilities of $D(R)$ and $D(R{\cal T})$. 

Then  the
operator (or representation
matrix) for $Q$ can be
defined as
\begin{equation}
P_{ll'}^{nn'}(Q)=\begin{cases}
\delta_{nn'}\tilde{\delta}_{\boldsymbol{d}_{l}^{n},Q\boldsymbol{d}_{l'}^{n'}}D^{n}(R) & \text{ Q does not contain \ensuremath{{\cal T}}}\\
\delta_{nn'}\tilde{\delta}_{\boldsymbol{d}_{l}^{n},Q\boldsymbol{d}_{l'}^{n'}}D^{n}(R{\cal T}) & \text{ Q contains \ensuremath{{\cal T}}}
\end{cases}\label{eq:pop}
\end{equation}
where$\tilde{\delta}_{\boldsymbol{d}_{l}^{n},Q\boldsymbol{d}_{l'}^{n'}}$
is equal to 1 only when
$d_{l}^{n}$ and $Qd_{l'}^{n'}$ differ by a lattice vector and to
0 otherwise (it can also be written as $\tilde{\delta}_{\boldsymbol{d}_{l}^{n},Q\boldsymbol{d}_{l'}^{n'}}=\delta_{\boldsymbol{d}_{l}^{n},Q\boldsymbol{d}_{l'}^{n'}+\boldsymbol{R}_{s}}$
if a suitable lattice vector $\boldsymbol{R}_{s}$ is choosen). The Hamiltonian under constraint of $Q$ can
be written as 
\begin{equation}
P(Q)^{-1}H(\boldsymbol{k})P(Q)=H(R^{-1}\boldsymbol{k})\label{eq:com}
\end{equation}
for $Q=\{R|\boldsymbol{t}\}$, and 
\begin{equation}
P(Q)^{-1}H(\boldsymbol{k})P(Q)=H^{*}(-R^{-1}\boldsymbol{k})\label{eq:comtri}
\end{equation}
for $Q=\{R|\boldsymbol{t}\}{\cal T}$ {[}See
~\ref{app} for proof of Eqs.(\ref{eq:hopp}--\ref{eq:hopptri})
and Eqs.(\ref{eq:com}--\ref{eq:comtri})).
Eqs.(\ref{eq:com}--\ref{eq:comtri})
are key point to generate the symmetry adopted tight-binding model.
In MagneticTB we first generate the Hamiltonian with only translation
symmetry and then use Eqs.(\ref{eq:com}--\ref{eq:comtri})
to the simplify the
Hamiltonian (see Fig.~\ref{fig:fig1}(b))
for the workflow of MagneticTB{]}.
The database for tight-binding model of 1651 magnetic space group
can be found in our later work \citep{zhang_database}. 


\section{Capabilities of MagneticTB}

\label{sec:Capabilities}

\subsection{Installation}

To install the MagneticTB, unzip the "MagneticTB.zip" file and copy
the MagneticTB directory to any of
the following four paths:
\begin{itemize}
\item  \lstinline!FileNameJoin[{$UserBaseDirectory, "Applications"}]!
\item  \lstinline!FileNameJoin[{$BaseDirectory, "Applications"}]!
\item  \lstinline!FileNameJoin[{$InstallationDirectory, "AddOns", "Packages"}]!
\item  \lstinline!FileNameJoin[{$InstallationDirectory, "AddOns", "Applications"}]!
\end{itemize}
Then one can use the package after running \lstinline!Needs["MagneticTB`"]!.
The version of Mathematica should higher or equal to 11.0. 

\subsection{Running}

\subsubsection{Core module}

To initialize the program one should identify the magnetic space group
and the orbital information in each Wyckoff positions. Here we provide
a function \lstinline!msgop! to show the symmetry information of
an arbitrary magnetic space group. The only input of the \lstinline!msgop!
is the magnetic space group number, one can get the symmetry information
by the following code: 


\begin{lstlisting}[backgroundcolor={\color{yellow!5!white}},mathescape=true]
msgop[gray[191]]
msgop[bnsdict[{191, 236}]]
msgop[ogdict[{191, 8, 1470}]]
Magnetic space group (BNS): {191.236,P6'/mm'm}
Primitive Lattice vactor: {{a,0,0},{-a/2,(Sqrt[3] a)/2,0},{0,0,c}}
Conventional Lattice vactor: {{a,0,0},{-a/2,(Sqrt[3] a)/2,0},{0,0,c}}
{{"1",{{1,0,0},{0,1,0},{0,0,1}},{0,0,0},F},
{"3z",{{0,-1,0},{1,-1,0},{0,0,1}},{0,0,0},F},
{"3z-1",{{-1,1,0},{-1,0,0},{0,0,1}},{0,0,0},F},
{"2x",{{1,-1,0},{0,-1,0},{0,0,-1}},{0,0,0},F},
...
\end{lstlisting}
here \lstinline!gray[191]! return the magnetic space group code of
gray space group 191, \lstinline!bnsdict[{191,236}]! return the magnetic
space group code of BNS No.~191.236, \lstinline!ogdict[{191,8,1470}]!
return the magnetic space group code of OG No.~191.8.1470. Then the
\lstinline!msgop! will print the standard lattice vector and the
symmetry operations
(for primitive cell) of the corresponding magnetic space group, which
can be the input of the \lstinline!init! function. Notice the magnetic
space group code is build-in constant in MagneticTB, users should
use \lstinline!gray, bnsdict, ogdict! functions rather than inputting
the magnetic space group code directly.

Then one can feed the above information to \lstinline!init! function
then the basic results of input structure can be generated. The \lstinline!init!
function have five mandatory
options neamly, \lstinline!lattice!, \lstinline!lattpar!, \lstinline!wyckoffposition!,
\lstinline!symminformation! and \lstinline!basisFunctions! (in ordinary
Wolfram language the options for functions are optional, but such
five options must be specified in MagneticTB in order to make the
input clear). The \lstinline!lattice! is the lattice vector of magnetic
system which can contain parameters, \lstinline!lattpar! is the parameters
in lattice vector to determine the bond length of magnetic system,and \lstinline!wyckoffposition! is a list to designate atomic position and magnetization-direction for each
Wyckoff positions in
the magnetic system. The
format of \lstinline!wyckoffposition! is: 
\[
\{\{\boldsymbol{a}_{1},\boldsymbol{m}_{1}\},\{\boldsymbol{a}_{2},\boldsymbol{m}_{2}\},...\}
\]
where $\boldsymbol{a}_{i}$ and $\boldsymbol{m}_{i}$ represent one
of the atomic positions and its
magnetization-directions
of the $i$-th Wyckoff
position, respectively. The \lstinline!symminformation! contain the
elements of the coset of magnetic space group with respect to translation
group, which can direct use the output of \lstinline!msgop! (notice
that the output of \lstinline!msgop! is standard symmetry operation
from
ISOTROPY \citep{litvin_1-_nodate,noauthor_isotropy_nodate}. However,
users can also use the non-standard structure as input, not limit
to the output of \lstinline!msgop!). The format of \lstinline!symminformation!
is: 
\[
\{\{n_{1},R_{1},t_{1},A_{1}\},\{n_{2},R_{2},t_{2},A_{2}\},...\}
\]
where $n_{i}$ is the name of symmetry operation,
$R_{i}$ and $t_{i}$ are the rotation and translation
part of symmetry operation,
and $A_{i}$ represents whether the symmetry operation
is combined with time reversal symmetry (\lstinline!"T"! for true
and \lstinline!"F"! for false). Finally, the \lstinline!basisFunctions!
is the basis function for each Wyckoff position, The format of \lstinline!basisFunctions!
is: 
\[
\{b_{1},b_{2},...\}
\]
where $b_{i}$ is the list
of basis functions of
the $i$-th Wyckoff position.
The build-in basis functions for spinless
case in MagneticTB is shown
in Table~\ref{tab:bf}. 
\begin{table}[ht]
\centering \caption{String codes representing basis functions and available values for
\lstinline!basisFunctions!}
\label{tab:bf} %
\begin{tabular}{cc|cc}
\hline 
Basis function  & String  & Basis function  & String \tabularnewline
\hline 
$s$  & \lstinline!"s"! & $p_{x}$  & \lstinline!"px"! \tabularnewline
\hline 
$p_{y}$  & \lstinline!"py"! & $p_{z}$  & \lstinline!"pz"! \tabularnewline
\hline 
$p_{x}+ip_{y}$  & \lstinline!"px+ipy"! & $p_{x}-ip_{y}$  & \lstinline!"px-ipy"! \tabularnewline
\hline 
$d_{z^{2}}$  & \lstinline!"dz2"! & $d_{xy}$  & \lstinline!"dxy"! \tabularnewline
\hline 
$d_{yz}$  & \lstinline!"dyz"! & $d_{xz}$  & \lstinline!"dxz"! \tabularnewline
\hline 
$d_{x^{2}-y^{2}}$  & \lstinline!"dx2-y2"!  &  & \tabularnewline
\hline 
\end{tabular}
\end{table}

When  spin is considered,
for build-in basis functions, add \lstinline!"up"! or \lstinline!"dn"!
after basis functions string, e.g. for $\ket{p_{x}\uparrow}$, the
basis function for spin-up case is \lstinline!"pxup"!. However, users
may use other basis functions such as $f$, $\ket{3/2,1/2}$ orbitals.
In such cases, users can directly input the analytical expression
of basis functions. For example, if only consider $f_{xyz}$ orbital,
one should input \lstinline!basisFunctions -> {{x*y*z}}]!. The analytical
expressions of basis functions can be simply obtained from quantum
mechanics or group theory books \citep{bradley_mathematical_1972,dresselhaus_group_2008,hergert_group_2018}.
\begin{lstlisting}[backgroundcolor={\color{yellow!5!white}},mathescape=true]
sgop=msgop[gray[191]];
init[
lattice -> {{a,0,0},{-(a/2),(Sqrt[3] a)/2,0},{0,0,c}},
lattpar -> {a -> 1, c -> 3},
wyckoffposition -> {{{1/3, 2/3, 0}, {0, 0, 0}}},
symminformation -> sgop,
basisFunctions -> {{"pz"}}];
\end{lstlisting}
\begin{table}[ht]
\centering \caption{Basic results of \lstinline!init!}
\label{tab:pro} %
\begin{tabular}{c|c}
\hline 
properties  & illustrate of properties \tabularnewline
\hline 
\lstinline!atompos!  & atomic position and magnetization-direction for each atom \tabularnewline
\hline 
\lstinline!wcc!  & $\boldsymbol{d}_{l}$ for each basis function \tabularnewline
\hline 
\lstinline!reclatt!  & reciprocal lattice vector for given structure \tabularnewline
\hline 
\lstinline!symmetryops! & $P_{Q}$ for each symmetry operation \tabularnewline
\hline 
\lstinline!unsymham!  & generate the  Hamiltonian with only translation symmetry \tabularnewline
\hline
\lstinline!symmcompile!  & summary of $P(Q)$, see main text for detail \tabularnewline
\hline 
\lstinline!bondclassify!  & summary of bonds information, see main text for detail\tabularnewline
\hline 
\end{tabular}
\end{table}

After inputting the above
five options appropriately,
one can run \lstinline!init! and obtain the basic results. Here we
introduce two important basic results: \lstinline!symmcompile! and
\lstinline!bondclassify!, the other properties are given in Table.~\ref{tab:pro}.
The format of \lstinline!symmcompile! is 
\[
\{\{N_{1},\{n_{1},R_{1},t_{1},A_{1}\},P_{1},R_{1}^{k}\},\{N_{2},\{n_{2},R_{2},t_{2},A_{2}\},P_{2},R_{2}^{k}\},...\}
\]
where $N_{i}$, $P_{i}$, $R_{i}^{k}$ are the label, the
symmetry operator (Eq.(\ref{eq:pop})) and the rotation acting
on $\boldsymbol{k}$ space of the
$i$-th symmetry operation,
respectively. For example 
\begin{lstlisting}[backgroundcolor={\color{yellow!5!white}},mathescape=true]
symmcompile
{{1,{"1",{{1,0,0},{0,1,0},{0,0,1}},{0,0,0},F},{{1,0},{0,1}},
		{{1,0,0},{0,1,0},{0,0,1}}},
{2,{"6z",{{1,-1,0},{1,0,0},{0,0,1}},{0,0,0},F},{{0,1},{1,0}},
	{{0,-1,0},{1,1,0},{0,0,1}}},
...}
\end{lstlisting}
The format of \lstinline!bondclassify! is 
\[
\{\{l_{i},n{}_{i},\{\{p_{i,j},p_{i,k}\},...\}\},...\}
\]
where $l_{i}$ is the bond length of the $(i-1)$-th neighbour hopping
($i=1$ for on-site hopping), $n_{i}$ is the number of
the $(i-1)$-th neighbour's bonds, $\{p_{i,j},p_{i,k}\}$ is the
atomic position of $n_{i}$ corresponding bonds.

Till this moment, symmetry adopted tight-binding model for magnetic
system is ready to be generated. By using the \lstinline!symham[n]!
function, one can obtain the symmetry adopted tight-binding model.
When $n=1$, \lstinline!symham[1]! return the Hamiltonian with only
on-site hopping, $n=2$ return the Hamiltonian with only
nearest-neighbour hopping and so on. By default, MagneticTB will
check all the input symmetry operations
to ensure the Hamiltonian is correct. However, it may last long
time when the structure is complex. In principle, only the generators
of the (magnetic) space
group are
enough to get the Hamiltonian. Therefore, one can specify the symmetry
operations
by \lstinline!symmetryset->list! in \lstinline!symham! where \lstinline!list! is the list of
indexes of the symmetry operations.
For example, \lstinline!symham[2,symmetryset->{2}]! will generate
the Hamiltonian with only $C_{6z}$ symmetry for nearest-neighbour
hopping. \lstinline!symmetryset! can not only save
computing resource but can also investigate the Hamiltonian for symmetry
breaking cases. The parameters for each neighbour in MagneticTB are
given in Table.~{\ref{tab:para}}. 
\begin{table}[ht]
\centering \caption{String codes representing $n$-th neighbour hoppings for \lstinline!symham!}
\label{tab:para} %
\begin{tabular}{c|c|c|c|c}
\hline 
On-site energy  & Nearest  & Second-nearest  & Third-nearest  & $(k-1)$-th nearest \tabularnewline
\hline 
\lstinline!e1,e2,..! & \lstinline!t1,t2,...! & \lstinline!r1,r2,...!  & \lstinline!s1,s2,...!  & \lstinline!pkn1,pkn2,...! \tabularnewline
\hline 
\end{tabular}
\end{table}

\subsubsection{Plot module}

After tight-binding model being generated, there may exist many parameters,
one can use \lstinline!bandManipulate! function to manipulate the
band structure to investigate the relationship between band structure
and parameters. The format of \lstinline!bandManipulate! is 
\begin{lstlisting}[backgroundcolor={\color{yellow!5!white}},mathescape=true]
bandManipulate[{{{k1,k2},{name of k1, name of k2}},...},np,Hamiltonian]
\end{lstlisting}
where \lstinline!np! is the number of $k$ points per line. Then
one can easily check the band structure of different parameters. When
the proper parameters are obtained, one can use \lstinline!bandplot!
to plot the band structure 
\begin{lstlisting}[backgroundcolor={\color{yellow!5!white}},mathescape=true]
bandplot[{{{k1,k2},{name of k1, name of k2}},...},np,Hamiltonian,parameters]
\end{lstlisting}
see section.~\ref{sec:example} for concrete example. 


\subsubsection{IO module}

\label{sec:IO} In MagnetTB, one can get the tight-binding model for
magnetic system. 
However, MagneticTB 
do not calculate
the other properties
(such as surface states, finding the gap-less point and so on)
directly, since it will generally cost too much computing resources.
It is better to do such heavy calculations by Fortran, Python or C.
Therefore we develop \lstinline!hopp! function to convert the symmetry
adopted tight-binding model to "wannier90\_hr.dat" format, which
is convenient to interface with WannierTools \citep{wu_wanniertools:_2018},
Z2Pack \citep{gresch_z2pack:_2017}, PythTB \citep{yusufaly_tight-binding_nodate}
and our home-made
package Wannflow \citep{zhang_high-throughput_2018,li_high_2018}.
The "wannier90\_hr.dat" in Wannier90 use the following convention
\citep{mostofi_wannier90:_2008}, namely conventions II: 
\begin{equation}
\begin{split}\tilde{\psi}_{lm\boldsymbol{k}}^{n}(\boldsymbol{r}) & =\frac{1}{\sqrt{N}}\sum_{\boldsymbol{R}_{j}}e^{i\boldsymbol{k}\cdot\boldsymbol{R}_{j}}\varphi_{lm}^{n}(\boldsymbol{r}-\boldsymbol{R}_{j}-\boldsymbol{d}_{l}^{n})\\
\tilde{H}_{lml'm'}^{nn'}(\boldsymbol{k}) & =\sum_{\boldsymbol{R}_{j}}e^{i\boldsymbol{k}\cdot\boldsymbol{R}_{j}}E_{mm'}(\boldsymbol{d}_{l}^{n},\boldsymbol{R}_{j}+\boldsymbol{d}_{l'}^{n'})
\end{split}
\end{equation}
which is different form MagneticTB in Eq.~(\ref{eq:covI}), the relationship
between two conventions is 
\begin{equation}
\begin{split}\tilde{H}(\boldsymbol{k})= & V(\boldsymbol{k})H(\boldsymbol{k})V^{\dagger}(\boldsymbol{k})\\
V_{ll'}^{nn'}(\boldsymbol{k})= & e^{i\boldsymbol{k}\cdot\boldsymbol{d}_{l}^{n}}\delta_{ll'}\delta_{nn'}
\end{split}
\label{conII}
\end{equation}
In MagneticTB (convention I) the operation
matrix defined in Eq.~(\ref{eq:pop}) is $\boldsymbol{k}$ independent
while the Hamiltonian is non-periodic by shifting the reciprocal vector
$\boldsymbol{G}$ 
\begin{equation}
\begin{split}H(\boldsymbol{k}+\boldsymbol{G})= & V^{\dagger}(\boldsymbol{G})H(\boldsymbol{k})V(\boldsymbol{G})\end{split}
\end{equation}
By contrast, in conventions II the Hamiltonian is periodic. i.e 
$\tilde{H}(\boldsymbol{k}+\boldsymbol{G})=\tilde{H}(\boldsymbol{k})$.
The format of \lstinline!hopp! function is 
\begin{lstlisting}[backgroundcolor={\color{yellow!5!white}},mathescape=true]
hopp[Hamiltonian,parameters]
\end{lstlisting}
See
section.~\ref{sec:example} for concrete example. One can also use
\lstinline!symmhamII[ham]! to generate the Wolfram expression for
Hamiltonian in convention II. Notice that
\lstinline!hopp! function (but not \lstinline!symmhamII!) is only
applied
to the output of \lstinline!symham! function, and
that expressions explicitly including
\lstinline!Sin! or \lstinline!Cos! may not work well.
Be
careful to use it. 

\section{Examples}

\label{sec:example} 

\subsection{Three-band tight-binding model for MoS$_{2}$}

MoS$_2$ monolayer has direct bandgap in the visible range, strong spin-orbit coupling, and
rich valley related physics, which make it an candidate for nanoelectronic, optoelectronic, and
valleytronic applications
 \citep{wang_electronics_2012,Liu_Yao_2015_44_2643__Electronic}.
The space group of MoS$_{2}$ is $P\overline{6}m2$ (space group No.~187).
Considering the Mo atom
 at $1a$ Wyckoff
position and
using the $d_{z^{2}}$, $d_{xy}$, and $d_{x^{2}-y^{2}}$ orbitals,
the
model can be obtained by 
\begin{lstlisting}[backgroundcolor={\color{yellow!5!white}},mathescape=true]
sgop = msgop[gray[187]];
tran = {{1, -1, 0}, {0, 1, 0}, {0, 0, 1}};
sgoptr = MapAt[FullSimplify[tran.#.Inverse@tran] &, sgop, {;; , 2}];
init[
  lattice -> {{1, 0, 0}, {1/2, Sqrt[3]/2, 0}, {0, 0, 10}},
  lattpar -> {},
  wyckoffposition -> {{{0, 0, 0}, {0, 0, 0}}},
  symminformation -> sgoptr,
  basisFunctions -> {{"dz2", "dxy", "dx2-y2"}}];
mos2 = Sum[symham[i, symmetryset -> {9, 11, 13}], {i, {1, 2}}]; 
mos2Liu = mos2 /. Thread[{kx, ky, kz} -> ({kx, ky, kz} (2 Pi)).Inverse@reclatt];
\end{lstlisting}
This
gives exactly the same
results as in Ref.~\citep{liu_three-band_2013}. The relationship
of parameters between Ref.~\citep{liu_three-band_2013} and MagneticTB
are 
\begin{center}
\begin{tabular}{c|cccccccc}
\hline 
Ref.~\citep{liu_three-band_2013}  & $\epsilon_{1}$  & $\epsilon_{2}$  & $t_{0}$  & $t_{1}$  & $t_{2}$  & $t_{11}$  & $t_{12}$  & $t_{22}$ \tabularnewline
\hline 
MagneticTB  & \lstinline!e1! & \lstinline!e2!  & \lstinline!t1!  & \lstinline!t2! & \lstinline!t4!  & \lstinline!t3! & \lstinline!t5!  & \lstinline!t6!\tabularnewline
\hline 
\end{tabular}
\par\end{center}

\subsection{Graphene}

Graphene with linear dispersion around Fermi level is one of the
most important materials
in spintronics \citep{castro_neto_electronic_2009}. The magnetic
space group of graphene is $P6/mmm1'$ (BNS No. 191.234).
There
are two C atoms 
at $2c$ Wyckoff position, and the bands
near Fermi energy are
mainly from $p_{z}$ orbital. The above information is enough to establish
the tight-binding model near Fermi energy of graphene. The model can
be obtained as follow 
\begin{lstlisting}[backgroundcolor={\color{yellow!5!white}},mathescape=true]
Needs["MagneticTB`"]
sgop = msgop[gray[191]];
init[
lattice ->  {{a,0,0},{-(a/2),(Sqrt[3] a)/2,0},{0,0,c}},
lattpar -> {a -> 1, c -> 3},
wyckoffposition -> {{{1/3, 2/3, 0}, {0, 0, 0}}},
symminformation -> sgop,
basisFunctions -> {{"pz"}}];
ham = Sum[symham[i], {i, 3}]; MatrixForm[ham]
\end{lstlisting}
output: 
\[
\left[\begin{array}{cc}
e_{1}+2r_{1}(\cos(k_{x}+k_{y})+\cos k_{x}+\cos k_{y}) & t_{1}e^{i\left(-\frac{2k_{x}}{3}-\frac{k_{y}}{3}\right)}+t_{1}e^{i\left(\frac{k_{x}}{3}-\frac{k_{y}}{3}\right)}+t_{1}e^{i\left(\frac{k_{x}}{3}+\frac{2k_{y}}{3}\right)}\\
\dagger & e_{1}+2r_{1}(\cos(k_{x}+k_{y})+\cos k_{x}+\cos k_{y})
\end{array}\right]
\]
For
spin-orbital coupling
case, the only thing which needs to change is the
basis functions 
\begin{lstlisting}[backgroundcolor={\color{yellow!5!white}},mathescape=true]
basisFunctions -> {{"pzup", "pzdn"}}
\end{lstlisting}
and the corresponding Hamiltonian reads 
\[
\left[\begin{array}{cccc}
e_{1}+h^{-} & 0 & h' & 0\\
0 & e_{1}+h^{+} & 0 & h'\\
\dagger & 0 & e_{1}+h^{+} & 0\\
0 & \dagger & 0 & e_{1}+h^{-}
\end{array}\right]
\]
where $h^{\pm}=\pm2r_{1}(\sin k_{x}+\sin k_{y}-\sin k_{x}+k_{y})+2r_{2}(\cos(k_{x}+k_{y})+\cos k_{x}+\cos k_{y})$,
$h'=t_{1}e^{i\left(-\frac{2k_{x}}{3}-\frac{k_{y}}{3}\right)}+t_{1}e^{i\left(\frac{k_{x}}{3}-\frac{k_{y}}{3}\right)}+t_{1}e^{i\left(\frac{k_{x}}{3}+\frac{2k_{y}}{3}\right)}$.
Such model is corresponding to the first $Z_{2}$ topological insulator
\citep{kane_z2_2005}. After get the Hamiltonian, we can use \lstinline!bandManipulate!
and \lstinline!bandplot! to plot the band structure, and click the
\lstinline!"ExportData"! button to print the value of parameters.
\begin{lstlisting}[backgroundcolor={\color{yellow!5!white}},mathescape=true]
path={
{{{0,0,0},{0,1/2,0}},{"\[CapitalGamma]","M"}},
{{{0,1/2,0},{1/3,1/3,0}},{"M","K"}},
{{{1/3,1/3,0},{0,0,0}},{"K","\[CapitalGamma]"}}
};
bandManipulate[path, 20, ham]
bandManipulate[path, 20, hamsoc]
bandplot[path, 200, ham, {e1 -> 0, t1 -> 0.5, r1-> 0}]
bandplot[path, 200, hamsoc, {e1 -> 0, t1 -> 0.5, r1-> 0.02, r2->0}]
\end{lstlisting}
\begin{figure}[h]
\includegraphics[width=0.8\textwidth]{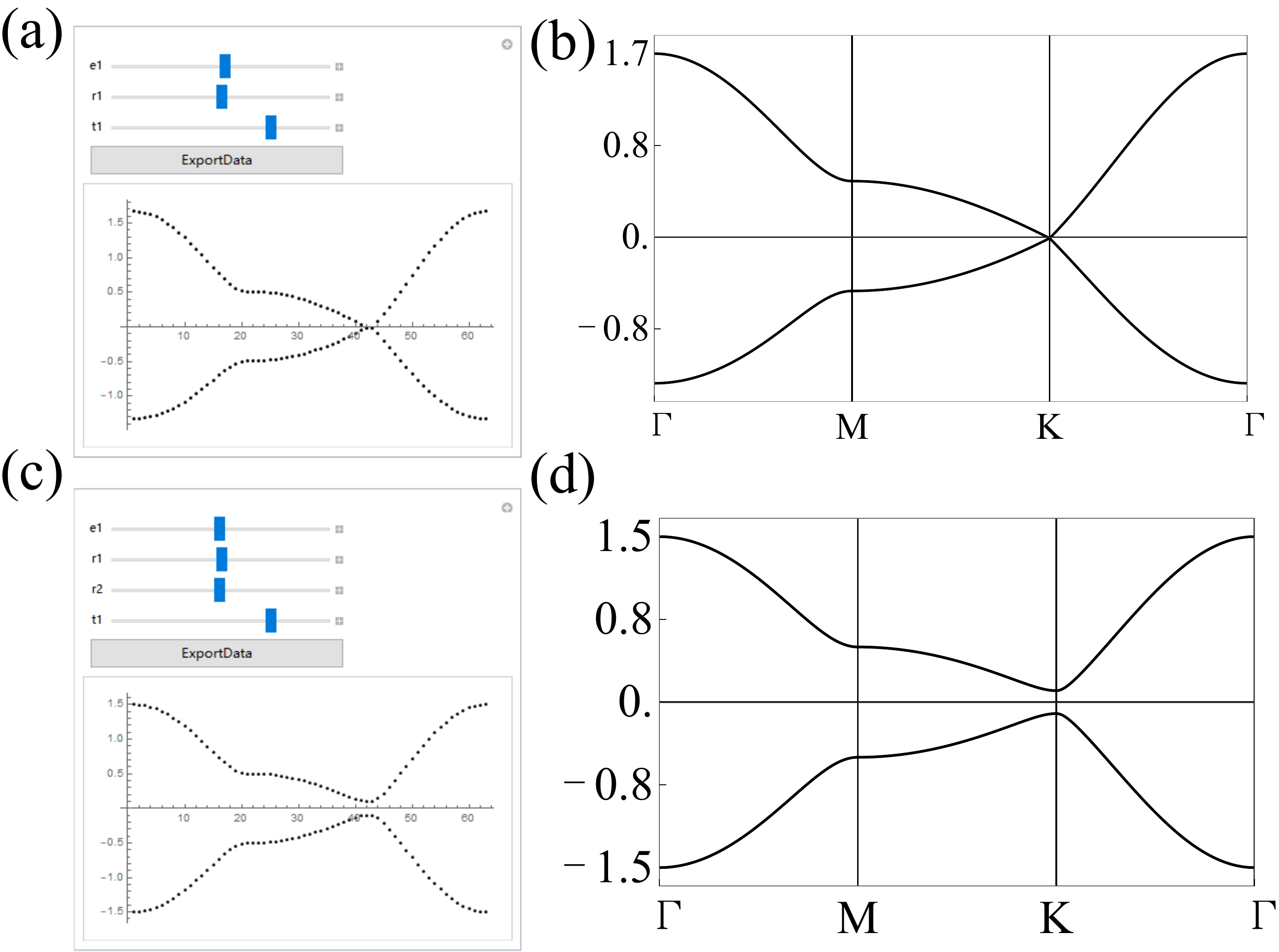} \caption{Output of \lstinline!bandManipulate! and \lstinline!bandplot! for
graphene without considering spin (a-b) and with spin (c-d).}
\label{fig:graphene} 
\end{figure}

Moreover we can get the wannier90\_hr.dat by \lstinline!hop! function
for this model. 
\begin{lstlisting}[backgroundcolor={\color{yellow!5!white}},mathescape=true]
hop[hamsoc, {e1 -> 0, r1 -> 0.02, r2 -> 0, t1 -> 0.5}]
Generated by MagneticTB
4
7
1    1    1    1    1    1    1
-1    -1     0     1     1    0.00000000    0.02000000
-1    -1     0     2     1    0.00000000    0.00000000
-1    -1     0     3     1    0.00000000    0.00000000
-1    -1     0     4     1    0.00000000    0.00000000
-1    -1     0     1     2    0.00000000    0.00000000
-1    -1     0     2     2    0.00000000   -0.02000000
...
\end{lstlisting}

\subsection{Magnetic C-3 Weyl point}

\begin{figure}[h]
\includegraphics[width=1\textwidth]{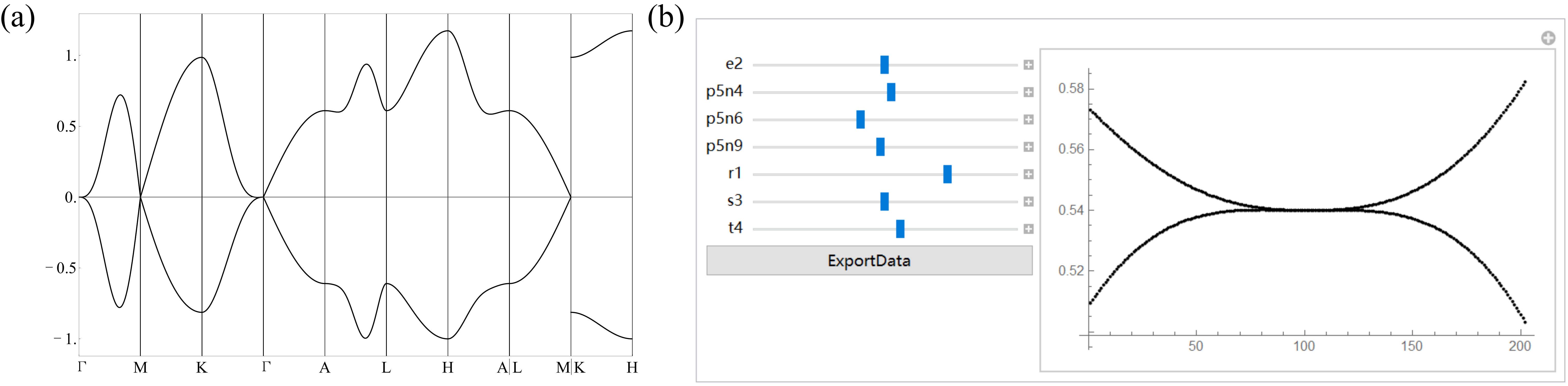} \caption{(a) Output of \lstinline!bandplot! for magnetic C-3 Weyl point, (b)
Output of \lstinline!bandManipulate! for magnetic cubic nodal-line.}
\label{fig:c4cnl} 
\end{figure}

The charge-3 (C-3) Weyl point is a 0D two-fold band degeneracy with
Chern number $|C|=3$. Encyclopedia of emergent particles tell us
that the C-3 Weyl point always appear at least in a pair or coexist
with nodal surface in nonmagnetic systems \citep{yu_encyclopedia_2020}.
Here we confirm that in magnetic system, due to the breaking of time
reversal symmetry ${\cal T}$, the C-3 Weyl point can uniquely coexist
with conventional Weyl points. Consider the type IV magnetic space
group $P_{c}3$ (BNS No. 143.3).
The
generator of the group
is $C_{3z}$ and $\{E|00\frac{1}{2}\}{\cal T}$,
Put
$\ket{p_{x}+ip_{y}\uparrow},\ket{p_{x}-ip_{y}\downarrow}$ basis functions
at
Wyckoff position $2a$, and
then the symmetry operator for $C_{3z}$ and $E\{00\frac{1}{2}\}{\cal T}$
are 
\[
\begin{split}C_{3z} & =-\sigma_{z}\\
\{E|00\frac{1}{2}\}{\cal T} & =i\sigma_{y}
\end{split}
\]
Under
this bases
the effective Hamiltonian
at $\Gamma$ point can be written as 
\begin{eqnarray}
H_{\text{C-3\ WP}} & = & \epsilon+\alpha k_{z}\sigma_{x}+ck_{\parallel}^{2}+\beta(k_{x}+e^{-i\frac{\pi}{3}}k_{y})^{3}\sigma_{z}+h.c.\label{eq:c3w}
\end{eqnarray}
where $\epsilon,c$ are real parameters and $\alpha,\beta$ are complex
parameters. Besides, there
are another three essential Weyl points locate at $(\pi,0,0),(0,\pi,0),(\pi,\pi,0)$.
Since the $C_{3z}$ symmetry does not change the Chern number of Weyl
points, the Chern number 
at $M$ has to be $\pm1$. According to no-go theorem, the Chern number
of $\Gamma$ is $\mp3$. One can easily check that the Chern number
of Eq.~(\ref{eq:c3w}) is $\pm3$. The degeneracies of $\Gamma$
and $M$ are because $(\{E|00\frac{1}{2}\}{\cal T})^{2}=-1$ at $(0/\pi,0/\pi,0)$. The model can be obtained as follow
\begin{lstlisting}[backgroundcolor={\color{yellow!5!white}},mathescape=true]
sgop = msgop[bnsdict[{143, 3}]];
init[lattice -> {{Sqrt[3]/2, -( 1/2), 0}, {0, 1, 0}, {0, 0, 2}},
lattpar -> {},
wyckoffposition -> {{{0, 0, 0}, {0, 0, 1}}},
symminformation -> sgop,
basisFunctions -> {{{x + I y, 0}, {0, x - I y}}}];
c3w = Sum[symham[i], {i, {2, 4}}];
c3w2band = Table[c3w[[i, j]], {i, {1, 4}}, {j, {1, 4}}];
\end{lstlisting}
The
band structure of \lstinline!c3w2band! is shown
in Fig.~\ref{fig:c4cnl}(a).

\subsection{Magnetic cubic nodal-line}

Topological high order nodal line is that the energy difference between
the bands are non-linear,and the order of energy dispersion around the degeneracy points plays
an important role in
different physical properties, such as density of states, Berry phase
and Landau-level \citep{yu_quadratic_2019}.
Recently, Zhang et. al. proposed high order nodal-line in magnetic
system \citep{zhang_magnetic_nodate}. In this example, we use MagneticTB
to generate magnetic cubic nodal-line. Generally, the magnetic cubic
nodal-line is protected by $C_{6z}$ and $M_{x}$ symmetries,
there are many magnetic space groups
which contains
the above two symmetries. Consider the type IV magnetic space group
$P_{c}6cc$ (BNS No. 184.196), and put $\ket{p_{x}+ip_{y}\uparrow},\ket{p_{x}-ip_{y}\downarrow}$
basis functions at
Wyckoff position $2a$.
Then the tight-binding
model can be generated by 
\begin{lstlisting}[backgroundcolor={\color{yellow!5!white}},mathescape=true]
sgop = msgop[bnsdict[{184, 196}]];
init[lattice -> {{Sqrt[3]/2, -1/2, 0}, {0, 1, 0}, {0, 0, 2}},
lattpar -> {},
wyckoffposition -> {{{0, 0, 0}, {0, 0, 1}}},
symminformation -> sgop,
basisFunctions -> {{{x + I y, 0}, {0, x - I y}}}];
cnl = Sum[symham[i, symmetryset -> {2, 7, 13}], {i, 1, 5}];
cnl2band = Table[cnl[[i, j]], {i, {1, 4}}, {j, {1, 4}}]; 
MatrixForm[cnl2band]
path = {{
{{0, 1/10, 1/4}, {0, 0, 1/4}}, {"Q", "P"}},
{{{0, 0, 1/4}, {-1/10, 1/10, 1/4}}, {"P", "Q"}}};
bandManipulate[path, 20, cnl2band]
\end{lstlisting}
One
can check that no mater how the parameters change, the dispersion
of arbitrary point along
$\Gamma$-$A$ on
$k_{x}$-$k_{y}$ plane
are non-linear, see Fig.~\ref{fig:c4cnl}(b) which is consistent with
Ref.~\citep{zhang_magnetic_nodate}. 

\section{Conclusion}

In conclusion, we have developed a software package to
generate the symmetry-adopted tight-binding model for arbitrary magnetic
space group. The input parameters for MagneticTB are clear and easy
to set and both spinless and spinful Hamiltonian can be generated
automatically. Besides, some useful functions such as manipulating
the band structure, interfacing with other software are implemented,
which can be used for further study on the magnetic systems. Moreover,
MagneticTB can not only be used to investigate physical properties
of electronic systems, but also be used to study photonics,
ultracold, acoustic and mechanical systems \citep{ozawa_topological_2019,cooper_topological_2019,ma_topological_2019}.
Finally, an exciting direction for future is to apply the magnetic
field for the tight-binding
model in MagneticTB \citep{peierls_zur_1933}.

\section*{Acknowledgments}

ZZ acknowledges  the support by the NSF of China (Grant No. 12004028), the China Postdoctoral Science Foundation (Grant
No. 2020M670106),  the Fundamental
Research Funds for the Central Universities (ZY2018). GBL acknowledges the
support by the National Key R\&D Program of China (Grant No. 2017YFB0701600).
YY acknowledges the support by the National Key R\&D Program of China (Grant No. 2020YFA0308800), the NSF of China (Grants Nos. 11734003, 
12061131002),  the Strategic Priority Research Program of Chinese Academy of Sciences (Grant No. XDB30000000).
\appendix

\section{\lyxadded{gbliu}{Sun May 09 20:09:32 2021}{}\lyxadded{gbliu}{Sun May 09 20:09:34 2021}{{}}}


\label{app} 
In this appendix, we use the translation operation $T(\boldsymbol{d})$.
For  $Q=\{R|\boldsymbol{v}\}$, we have
$QT(\boldsymbol{d})=\{R|\boldsymbol{v}\}\{E|\boldsymbol{d}\}=\{R|R\boldsymbol{d}+\boldsymbol{v}\}=T(Q\boldsymbol{d})R$. Thus
\begin{equation}
	\begin{split}E(\boldsymbol{d}_{j}^{n},(\boldsymbol{R}_{j}+\boldsymbol{d}_{l'}^{n'})) 
		&
		 =\braket{\hat{T}(\boldsymbol{d}_{l}^{n})\varPhi^{n}(\boldsymbol{r})|\hat{Q}^{\dagger}\hat{H}\hat{Q}|\hat{T}(\boldsymbol{d}_{l'}^{n'}+\boldsymbol{R}_{j})\varPhi^{n'}(\boldsymbol{r})}\\
		& =\braket{\hat{Q}\hat{T}(\boldsymbol{d}_{l}^{n})\varPhi^{n}(\boldsymbol{r})|\hat{H}|\hat{Q}\hat{T}(\boldsymbol{d}_{l'}^{n'}+\boldsymbol{R}_{j})\varPhi^{n'}(\boldsymbol{r})}\\
		& =\braket{\hat{T}(Q\boldsymbol{d}_{l}^{n})R\varPhi^{n}(\boldsymbol{r})|\hat{H}|\hat{T}(Q(\boldsymbol{d}_{l'}^{n'}+\boldsymbol{R}_{j}))R\varPhi^{n'}(\boldsymbol{r})}\\
		& =D^{n\dagger}(R)\braket{\hat{T}(Q\boldsymbol{d}_{l}^{n})\varPhi^{n}(\boldsymbol{r})|\hat{H}|\hat{T}(Q(\boldsymbol{d}_{l'}^{n'}+\boldsymbol{R}_{j}))\varPhi^{n'}(\boldsymbol{r})}D^{n'}(R)\\
		& =D^{n\dagger}(R)E(Q\boldsymbol{d}_{j}^{n},Q(\boldsymbol{R}_{j}+\boldsymbol{d}_{l'}^{n'}))D^{n'}(R)
	\end{split}
\end{equation}
which completes the proof of Eq.(\ref{eq:hopp}).

For  $Q=\{R|\boldsymbol{v}\}{\cal T}$, notice time reversal symmetry does not change the real space coordinates, i.e. ${\cal T}\boldsymbol{d}=\boldsymbol{d}$, we have
$QT(\boldsymbol{d})=\{R|\boldsymbol{v}\}{\cal T}\{E|\boldsymbol{d}\}=\{R|R\boldsymbol{d}+\boldsymbol{v}\}{\cal T}=T(Q\boldsymbol{d})R{\cal T}$. 
Use the fact that for anti-unitary operator $\hat{{\cal A}}$, $\braket{\hat{{\cal A}}\psi|\hat{{\cal A}}\phi}=\braket{\psi|\phi}^{*}$, 
then
\begin{equation}
	\begin{split}E^{*}(\boldsymbol{d}_{j}^{n},(\boldsymbol{R}_{j}+\boldsymbol{d}_{l'}^{n'}))
		& =\braket{\hat{Q}\hat{T}(\boldsymbol{d}_{l}^{n})\varPhi^{n}(\boldsymbol{r})|\hat{Q}\hat{H}|\hat{T}(\boldsymbol{d}_{l'}^{n'}+\boldsymbol{R}_{j})\varPhi^{n'}(\boldsymbol{r})}\\
		& =\braket{\hat{Q}\hat{T}(\boldsymbol{d}_{l}^{n})\varPhi^{n}(\boldsymbol{r})|\hat{H}|\hat{Q}\hat{T}(\boldsymbol{d}_{l'}^{n'}+\boldsymbol{R}_{j})\varPhi^{n'}(\boldsymbol{r})}\\
		& =\braket{\hat{T}(Q\boldsymbol{d}_{l}^{n})R{\cal T}\varPhi^{n}(\boldsymbol{r})|\hat{H}|\hat{T}(Q(\boldsymbol{d}_{l'}^{n'}+\boldsymbol{R}_{j}))R{\cal T}\varPhi^{n'}(\boldsymbol{r})}\\
		& =D^{n\dagger}(R{\cal T})\braket{\hat{T}(Q\boldsymbol{d}_{l}^{n})\varPhi^{n}(\boldsymbol{r})|\hat{H}|\hat{T}(Q(\boldsymbol{d}_{l'}^{n'}+\boldsymbol{R}_{j}))\varPhi^{n'}(\boldsymbol{r})}D^{n'}(R{\cal T})\\
		& =D^{n\dagger}(R{\cal T})E(Q\boldsymbol{d}_{j}^{n},Q(\boldsymbol{R}_{j}+\boldsymbol{d}_{l'}^{n'}))D^{n'}(R{\cal T})
	\end{split}
\end{equation}
which completes the proof of Eq.(\ref{eq:hopptri}).

Proof of Eq.(\ref{eq:com}), 
\begin{equation}
\begin{split}[H(\boldsymbol{k})P(Q)]_{ll'}^{nn'} & =\sum_{\mu\nu}H(\boldsymbol{k})_{l\mu}^{n\nu}P_{\mu l'}^{\nu n'}(Q)\\
 & =\sum_{\mu\nu}\sum_{\boldsymbol{R}_{j}}e^{i\boldsymbol{k}\cdot(\boldsymbol{R}_{j}+\boldsymbol{d}_{\mu}^{\nu}-\boldsymbol{d}_{l}^{n})}E(\boldsymbol{d}_{l}^{n},\boldsymbol{R}_{j}+\boldsymbol{d}_{\mu}^{\nu})\delta_{\nu n'}\delta_{\boldsymbol{d}_{\mu}^{\nu},Q\boldsymbol{d}_{l'}^{n'}+\boldsymbol{R}_{s}}D^{n'}(R)\\
 & =\sum_{\boldsymbol{R}_{j}}e^{i\boldsymbol{k}\cdot(\boldsymbol{R}_{j}+Q\boldsymbol{d}_{l'}^{n'}+\boldsymbol{R}_{s}-\boldsymbol{d}_{l}^{n})}E(\boldsymbol{d}_{l}^{n},\boldsymbol{R}_{j}+Q\boldsymbol{d}_{l'}^{n'}+\boldsymbol{R}_{s})D^{n'}(R)\\
 & \xlongequal{\boldsymbol{R}_{j}+\boldsymbol{R}_{s}\rightarrow R\boldsymbol{R}_{j}}\sum_{\boldsymbol{R}_{j}}e^{i\boldsymbol{k}\cdot R(\boldsymbol{R}_{j}+\boldsymbol{d}_{l'}^{n'}-Q^{-1}\boldsymbol{d}_{l}^{n})}E(\boldsymbol{d}_{l}^{n},Q(\boldsymbol{R}_{j}+\boldsymbol{d}_{l'}^{n'}))D^{n'}(R)\\
 & \xlongequal{\text{use Eq. (\ref{eq:hopp})}}\sum_{\boldsymbol{R}_{j}}e^{iR^{-1}\boldsymbol{k}\cdot(\boldsymbol{R}_{j}+\boldsymbol{d}_{l'}^{n'}-Q^{-1}\boldsymbol{d}_{l}^{n})}D^{n}(R)E(Q^{-1}\boldsymbol{d}_{l}^{n},\boldsymbol{R}_{j}+\boldsymbol{d}_{l'}^{n'})
\end{split}
\label{eq:HPR}
\end{equation}
\begin{equation}
\begin{split}[P(Q)H(R^{-1}\boldsymbol{k})]_{ll'}^{nn'} & =\sum_{\mu\nu}P_{l\mu}^{n\nu}(Q)H(R^{-1}\boldsymbol{k})_{\mu l'}^{\nu n'}\\
 & =\sum_{\mu\nu}\sum_{\boldsymbol{R}_{j}}\delta_{n\nu}\delta_{\boldsymbol{d}_{l}^{n},Q\boldsymbol{d}_{\mu}^{\nu}+\boldsymbol{R}_{s}}D^{n}(R)e^{iR^{-1}\boldsymbol{k}\cdot(\boldsymbol{R}_{j}+\boldsymbol{d}_{l'}^{n'}-\boldsymbol{d}_{\mu}^{\nu})}E(\boldsymbol{d}_{\mu}^{\nu},\boldsymbol{R}_{j}+\boldsymbol{d}_{l'}^{n'})\\
 & =\sum_{\boldsymbol{R}_{j}}D^{n}(R)e^{iR^{-1}\boldsymbol{k}\cdot(\boldsymbol{R}_{j}+\boldsymbol{d}_{l'}^{n'}-Q^{-1}(\boldsymbol{d}_{l}^{n}-\boldsymbol{R}_{s}))}E(Q^{-1}(\boldsymbol{d}_{l}^{n}-\boldsymbol{R}_{s}),\boldsymbol{R}_{j}+\boldsymbol{d}_{l'}^{n'})\\
 & =\sum_{\boldsymbol{R}_{j}}e^{iR^{-1}\boldsymbol{k}\cdot(\boldsymbol{R}_{j}+\boldsymbol{d}_{l'}^{n'}-Q^{-1}\boldsymbol{d}_{l}^{n}+R^{-1}\boldsymbol{R}_{s})}D^{n}(R)E(Q^{-1}\boldsymbol{d}_{l}^{n}-R^{-1}\boldsymbol{R}_{s},\boldsymbol{R}_{j}+\boldsymbol{d}_{l'}^{n'})\\
 & =\sum_{\boldsymbol{R}_{j}}e^{iR^{-1}\boldsymbol{k}\cdot(\boldsymbol{R}_{j}+\boldsymbol{d}_{l'}^{n'}-Q^{-1}\boldsymbol{d}_{l}^{n}+R^{-1}\boldsymbol{R}_{s})}D^{n}(R)E(Q^{-1}\boldsymbol{d}_{l}^{n},\boldsymbol{R}_{j}+\boldsymbol{d}_{l'}^{n'}+R^{-1}\boldsymbol{R}_{s})\\
 & \xlongequal{\boldsymbol{R}_{j}+R^{-1}\boldsymbol{R}_{s}\rightarrow\boldsymbol{R}_{j}}\sum_{\boldsymbol{R}_{j}}e^{iR^{-1}\boldsymbol{k}\cdot(\boldsymbol{R}_{j}+\boldsymbol{d}_{l'}^{n'}-Q^{-1}\boldsymbol{d}_{l}^{n})}D^{n}(R)E(Q^{-1}\boldsymbol{d}_{l}^{n},\boldsymbol{R}_{j}+\boldsymbol{d}_{l'}^{n'})
\end{split}
\label{eq:PHR-1k}
\end{equation}
In the above derivation we use the relation $Q^{-1}(\boldsymbol{d}_{l}^{n}-\boldsymbol{R}_{s})=Q^{-1}\boldsymbol{d}_{l}^{n}-R^{-1}\boldsymbol{R}_{s}$
($Q^{-1}$ is not linear).  Compare the last lines
of the above two equations and we can find they are equal to each
other, i.e.
\begin{equation}
[H(\boldsymbol{k})P(Q)]_{ll'}^{nn'}=[P(Q)H(R^{-1}\boldsymbol{k})]_{ll'}^{nn'}\ \ \ \ \ \Rightarrow\ \ \ \ \ H(\boldsymbol{k})P(Q)=P(Q)H(R^{-1}\boldsymbol{k})
\end{equation}

Proof of Eq.(\ref{eq:comtri}): similar
to Eq.(\ref{eq:HPR}) and Eq.(\ref{eq:PHR-1k}) we have
\begin{equation}
\begin{split}[H(\boldsymbol{k})P(Q)]_{ll'}^{nn'} & =\sum_{\mu\nu}H(\boldsymbol{k})_{l\mu}^{n\nu}P_{\mu l'}^{\nu n'}(Q)\\
 & =\sum_{\mu\nu}\sum_{\boldsymbol{R}_{j}}e^{i\boldsymbol{k}\cdot(\boldsymbol{R}_{j}+\boldsymbol{d}_{\mu}^{\nu}-\boldsymbol{d}_{l}^{n})}E(\boldsymbol{d}_{l}^{n},\boldsymbol{R}_{j}+\boldsymbol{d}_{\mu}^{\nu})\delta_{\nu n'}\delta_{\boldsymbol{d}_{\mu}^{\nu},Q\boldsymbol{d}_{l'}^{n'}+\boldsymbol{R}_{s}}D^{n'}(R\mathcal{T})\\
 & =\sum_{\boldsymbol{R}_{j}}e^{i\boldsymbol{k}\cdot(\boldsymbol{R}_{j}+Q\boldsymbol{d}_{l'}^{n'}+\boldsymbol{R}_{s}-\boldsymbol{d}_{l}^{n})}E(\boldsymbol{d}_{l}^{n},\boldsymbol{R}_{j}+Q\boldsymbol{d}_{l'}^{n'}+\boldsymbol{R}_{s})D^{n'}(R\mathcal{T})\\
 & \xlongequal{\boldsymbol{R}_{j}+\boldsymbol{R}_{s}\rightarrow R\boldsymbol{R}_{j}}\sum_{\boldsymbol{R}_{j}}e^{i\boldsymbol{k}\cdot R(\boldsymbol{R}_{j}+\boldsymbol{d}_{l'}^{n'}-Q^{-1}\boldsymbol{d}_{l}^{n})}E(\boldsymbol{d}_{l}^{n},Q(\boldsymbol{R}_{j}+\boldsymbol{d}_{l'}^{n'}))D^{n'}(R\mathcal{T})\\
 & \xlongequal{\text{use Eq. (\ref{eq:hopptri})}}\sum_{\boldsymbol{R}_{j}}e^{iR^{-1}\boldsymbol{k}\cdot(\boldsymbol{R}_{j}+\boldsymbol{d}_{l'}^{n'}-Q^{-1}\boldsymbol{d}_{l}^{n})}D^{n}(R\mathcal{T})E^{*}(Q^{-1}\boldsymbol{d}_{l}^{n},\boldsymbol{R}_{j}+\boldsymbol{d}_{l'}^{n'})
\end{split}
\end{equation}
\begin{equation}
\begin{split}[P(Q)H^{*}(-R^{-1}\boldsymbol{k})]_{ll'}^{nn'} & =\sum_{\mu\nu}P_{l\mu}^{n\nu}(Q)H^{*}(-R^{-1}\boldsymbol{k})_{\mu l'}^{\nu n'}\\
 & =\sum_{\mu\nu}\sum_{\boldsymbol{R}_{j}}\delta_{n\nu}\delta_{\boldsymbol{d}_{l}^{n},Q\boldsymbol{d}_{\mu}^{\nu}+\boldsymbol{R}_{s}}D^{n}(R\mathcal{T})e^{iR^{-1}\boldsymbol{k}\cdot(\boldsymbol{R}_{j}+\boldsymbol{d}_{l'}^{n'}-\boldsymbol{d}_{\mu}^{\nu})}E^{*}(\boldsymbol{d}_{\mu}^{\nu},\boldsymbol{R}_{j}+\boldsymbol{d}_{l'}^{n'})\\
 & =\sum_{\boldsymbol{R}_{j}}D^{n}(R\mathcal{T})e^{iR^{-1}\boldsymbol{k}\cdot(\boldsymbol{R}_{j}+\boldsymbol{d}_{l'}^{n'}-Q^{-1}(\boldsymbol{d}_{l}^{n}-\boldsymbol{R}_{s}))}E^{*}(Q^{-1}(\boldsymbol{d}_{l}^{n}-\boldsymbol{R}_{s}),\boldsymbol{R}_{j}+\boldsymbol{d}_{l'}^{n'})\\
 & =\sum_{\boldsymbol{R}_{j}}e^{iR^{-1}\boldsymbol{k}\cdot(\boldsymbol{R}_{j}+\boldsymbol{d}_{l'}^{n'}-Q^{-1}\boldsymbol{d}_{l}^{n}+R^{-1}\boldsymbol{R}_{s})}D^{n}(R\mathcal{T})E^{*}(Q^{-1}\boldsymbol{d}_{l}^{n}-R^{-1}\boldsymbol{R}_{s},\boldsymbol{R}_{j}+\boldsymbol{d}_{l'}^{n'})\\
 & =\sum_{\boldsymbol{R}_{j}}e^{iR^{-1}\boldsymbol{k}\cdot(\boldsymbol{R}_{j}+\boldsymbol{d}_{l'}^{n'}-Q^{-1}\boldsymbol{d}_{l}^{n}+R^{-1}\boldsymbol{R}_{s})}D^{n}(R\mathcal{T})E^{*}(Q^{-1}\boldsymbol{d}_{l}^{n},\boldsymbol{R}_{j}+\boldsymbol{d}_{l'}^{n'}+R^{-1}\boldsymbol{R}_{s})\\
 & \xlongequal{\boldsymbol{R}_{j}+R^{-1}\boldsymbol{R}_{s}\rightarrow\boldsymbol{R}_{j}}\sum_{\boldsymbol{R}_{j}}e^{iR^{-1}\boldsymbol{k}\cdot(\boldsymbol{R}_{j}+\boldsymbol{d}_{l'}^{n'}-Q^{-1}\boldsymbol{d}_{l}^{n})}D^{n}(R\mathcal{T})E^{*}(Q^{-1}\boldsymbol{d}_{l}^{n},\boldsymbol{R}_{j}+\boldsymbol{d}_{l'}^{n'})
\end{split}
\end{equation}
Compare the last lines of the above two equations and we can find
they are equal to each other, i.e.
\begin{equation}
[H(\boldsymbol{k})P(Q)]_{ll'}^{nn'}=[P(Q)H^{*}(-R^{-1}\boldsymbol{k})]_{ll'}^{nn'}\ \ \ \ \ \Rightarrow\ \ \ \ \ H(\boldsymbol{k})P(Q)=P(Q)H^{*}(-R^{-1}\boldsymbol{k})
\end{equation}

It is easy to verify that for $A,B$ 
not containing ${\cal T}$
and $C,D$ containing
${\cal T}$, $P(Q)$ obey the following
corepresentation algebra: 
\begin{equation}
\begin{split}P(A)P(B) & =P(AB)\\
P(A)P(C) & =P(AC)\\
P(C)P^{*}(A) & =P(CA)\\
P(C)P^{*}(D) & =P(CD)
\end{split}
\end{equation}

 \bibliographystyle{elsarticle-num-names}
\bibliography{magtb}


\end{document}